\documentstyle[12pt,epsfig,graphicx,amsmath]{article} 
\makeatletter \def\@cite#1#2{{[{#1}]\if@tempswa\typeout {IJCGA
warning: optional citation argument ignored: `#2'} \fi}}

\newcount\@tempcntc
\def\@citex[#1]#2{\if@filesw\immediate\write\@auxout{\string\citation{#2}}\fi
  \@tempcnta\z@\@tempcntb\m@ne\def\@citea{}\@cite{\@for\@citeb:=#2\do
    {\@ifundefined
       {b@\@citeb}{\@citeo\@tempcntb\m@ne\@citea\def\@citea{,}{\bf ?}\@warning
       {Citation `\@citeb' on page \thepage \space undefined}}%
    {\setbox\z@\hbox{\global\@tempcntc0\csname b@\@citeb\endcsname\relax}%
     \ifnum\@tempcntc=\z@ \@citeo\@tempcntb\m@ne
       \@citea\def\@citea{,}\hbox{\csname b@\@citeb\endcsname}%
     \else
     \advance\@tempcntb\@ne
      \ifnum\@tempcntb=\@tempcntc
      \else\advance\@tempcntb\m@ne\@citeo
      \@tempcnta\@tempcntc\@tempcntb\@tempcntc\fi\fi}}\@citeo}{#1}}
\def\@citeo{\ifnum\@tempcnta>\@tempcntb\else\@citea\def\@citea{,}%
  \ifnum\@tempcnta=\@tempcntb\the\@tempcnta\else
   {\advance\@tempcnta\@ne\ifnum\@tempcnta=\@tempcntb \else 
\def\@citea{--}\fi
    \advance\@tempcnta\m@ne\the\@tempcnta\@citea\the\@tempcntb}\fi\fi}
\makeatother

\def\boxit#1{\leavevmode\thinspace\hbox{\vrule\vtop{\vbox{\hrule%
        \vskip3pt\kern1pt\hbox{\vphantom{\bf/}\thinspace\thinspace%
        {\bf#1}\thinspace\thinspace}}\kern1pt\vskip3pt\hrule}\vrule}%
        \thinspace}
\def\Boxit#1{\noindent\vbox{\hrule\hbox{\vrule\kern3pt\vbox{
\advance\hsize-7pt\vskip-\parskip\kern3pt\bf#1 \hbox{\vrule height0pt
depth\dp\strutbox width0pt} \kern3pt}\kern3pt\vrule}\hrule}}

\newcommand{\gsim}{\lower.7ex\hbox{$\;\stackrel{\textstyle>}{\sim}\;$}}
\newcommand{\lsim}{\lower.7ex\hbox{$\;\stackrel{\textstyle<}{\sim}\;$}}
\newcommand{\be}{\begin{equation}} \newcommand{\ee}{\end{equation}}
\newcommand{\beq}{\begin{equation}} \newcommand{\eeq}{\end{equation}}
\newcommand{\bea}{\begin{eqnarray}} \newcommand{\eea}{\end{eqnarray}}

\def\baselinestretch{1}
\begin{document}
\catcode`@=11 \newtoks\@stequation
\def\subequations{\refstepcounter{equation}%
\edef\@savedequation{\the\c@equation}%
\@stequation=\expandafter{\theequation}
\edef\@savedtheequation{\the\@stequation}
\edef\oldtheequation{\theequation}
\def\theequation{\oldtheequation\alph{equation}}}
\def\endsubequations{\setcounter{equation}{\@savedequation}%
\@stequation=\expandafter{\@savedtheequation}%
\edef\theequation{\the\@stequation}\global\@ignoretrue

\noindent} \catcode`@=12
\begin{titlepage}

\title{\bf  Decaying Dark Matter \\and the PAMELA Anomaly}
\vskip3in \author{{\bf Alejandro Ibarra} and
{\bf David Tran\footnote{\baselineskip=16pt {\small E-mail addresses: {\tt
alejandro.ibarra@ph.tum.de, david.tran@ph.tum.de}}}}
\hspace{3cm}\\ \vspace{0.1cm}
{\normalsize\it  Physik-Department T30d, Technische Universit\"at M\"unchen,}\\[-0.05cm]
{\normalsize \it James-Franck-Stra\ss{}e, 85748 Garching, Germany.}
}  \date{}  \maketitle  \def\baselinestretch{1.15}
\begin{abstract}
\noindent 
Astrophysical and cosmological observations do not
require the dark matter particles to be absolutely stable.
If they are indeed unstable, their decay
into positrons might occur at a sufficiently large rate 
to allow the indirect detection of dark matter through 
an anomalous contribution to the cosmic positron flux.
In this paper we discuss the implications of the excess in the
positron fraction recently reported by the PAMELA collaboration
for the scenario of decaying dark matter.
To this end, we have performed a model-independent analysis
of possible signatures by studying various
decay channels in the case of both a fermionic and a scalar 
dark matter particle. We find that the steep rise
in the positron fraction measured by PAMELA at energies larger 
than 10 GeV can naturally be accommodated in several
realizations of the decaying dark matter scenario.

\end{abstract}

\thispagestyle{empty}
\vspace*{0.2cm} \leftline{November 2008} \leftline{}

\vskip-17.0cm \rightline{TUM-HEP 702/08}

\vskip3in

\end{titlepage}
\setcounter{footnote}{0} \setcounter{page}{1}
\newpage
\baselineskip=20pt

\noindent

\section{Introduction}

The spallation of primary cosmic-ray protons and other nuclei
on the interstellar medium produces a flux of secondary positrons 
which is expected to decrease monotonically with the 
energy~\cite{Moskalenko:1997gh}.
Interestingly, measurements of cosmic-ray positrons undertaken 
by a series of experiments over the last twenty years, 
HEAT~\cite{Barwick:1997ig}, CAPRICE~\cite{CAPRICE}, 
MASS~\cite{Grimani:2002yz} and AMS-01~\cite{Aguilar:2007yf},
indicated the existence of an excess of positrons at energies 
above 7 GeV with respect to the expectations from a purely secondary 
component. 
The discovery of this excess raised a lot of interest among the 
particle physics and astrophysics communities, which interpreted 
the excess as a possible indirect signature of dark matter.

If dark matter particles are weakly interacting, they
annihilate in the center of our Galaxy, producing
a primary flux of positrons. This possibility has been extensively
discussed over the last years as a potential explanation of the
HEAT anomaly, although typically large boost factors
have to be invoked in order to achieve sufficiently
high fluxes~\cite{annihilating,Baltz:1998xv,Hisano:2005ec}.

Nevertheless, dark matter self-annihilation is not 
the only possibility for 
the indirect detection of the dark matter. Strictly
speaking, the viability of a particle as a dark matter
candidate does not require its absolute stability,
but merely that the dark matter lifetime is longer than
the age of the Universe. Then, if the dark
matter decays proceed at a sufficiently high rate, the
decay products might be detectable. There are in fact some 
physically well-motivated dark matter candidates which
decay with very long lifetimes. For instance, gravitino dark matter
which is unstable due to a small breaking of $R$-parity constitutes a very
interesting scenario that leads to a thermal history of the
Universe consistent with the observed abundances of the primordial
elements, the observed dark matter relic abundance and
the observed baryon asymmetry~\cite{Buchmuller:2007ui}. 
The late decays of dark matter gravitinos produce 
a flux of gamma rays~\cite{Ibarra:2007wg,Ishiwata:2008cu}, 
positrons~\cite{Ibarra:2008qg,Ishiwata:2008cu},
antiprotons~\cite{Ibarra:2008qg} and neutrinos~\cite{Covi:2008jy} 
which contribute to the total fluxes received at the Earth. 
Remarkably, it has been pointed out that the EGRET 
anomaly in the extragalactic gamma-ray background~\cite{smr05} 
and the HEAT excess in the positron fraction~\cite{Barwick:1997ig} 
can be simultaneously explained by the decay of dark matter 
gravitinos with a mass $m\sim 150~{\rm GeV}$ and a lifetime 
$\tau\sim 10^{26}~{\rm s}$~\cite{Ibarra:2008qg,Ishiwata:2008cu}. 
Other candidates for decaying dark matter with electroweak masses include 
hidden gauge bosons~\cite{CTY}, hidden gauginos~\cite{Ibarra:2008kn}, 
right-handed sneutrinos in $R$-parity breaking scenarios~\cite{Chen:2008dh}
or baryonic bound states of messenger quarks~\cite{Hamaguchi:2008rv}. 

Very recently, the PAMELA collaboration~\cite{Picozza:2006nm}
has published measurements of the cosmic-ray positron fraction
performed with unprecedented accuracy~\cite{Adriani:2008zr}. 
These measurements have not only confirmed a significant deviation 
with respect to the expectations from a purely secondary component, 
but have also provided evidence for a 
very sharp rise of the spectrum at energies 7 -- 100 GeV. In view of 
the new results, it is important to study whether the scenario of decaying 
dark matter is consistent with the energy spectrum measured 
by PAMELA and what constraints the new data impose on the
nature of decaying dark matter. It should be borne in mind, though,
that the astrophysical uncertainties in the determination
of the secondary positron component are still large~\cite{Delahaye:2008ua}
and that nearby astrophysical sources such as pulsars might
produce sizable positron fluxes in the energy range explored
by PAMELA~\cite{pulsars}.

In order to keep the analysis as model-independent
as possible, we will analyze the cases that the dark matter
particle is either a fermion or a scalar, and we will compute the
predictions for the positron fraction for various decay channels
and different dark matter masses and lifetimes.
Namely, in the case of a fermionic dark matter particle $\psi$,
we will consider the two-body decay channels 
$\psi\rightarrow Z^0 \nu,\psi \rightarrow \;W^\pm \ell^\mp$, as
well as the three-body decay channels 
$\psi\rightarrow \ell^+ \ell^- \nu$, with $\ell=e,\;\mu,\;\tau$
being the charged leptons.
On the other hand, for a scalar dark matter particle $\phi$, we will
consider the two-body decay channels, 
$\phi \rightarrow Z^0 Z^0, \phi \rightarrow \; W^+ W^-,\;\phi \rightarrow \ell^+ \ell^-$.

This paper is organized as follows:
in Section 2 we will review the propagation of positrons in the
Galaxy. In Section 3 we will present our results for the positron 
fraction expected from the decay of a fermionic or a scalar dark matter
particle and we will discuss the sensitivity of the results to the 
choice of the propagation model. Lastly, in Section 4 we will present 
our conclusions.

\section{Positron Propagation}

Positron propagation in the Milky Way is commonly described by
a stationary two-zone diffusion model with cylindrical boundary 
conditions~\cite{ACR}. Under this approximation, 
the number density of positrons
per unit energy, $f_{e^+}(E,\vec{r},t)$, satisfies the following
transport equation:
\begin{equation}
0=\frac{\partial f_{e^+}}{\partial t}=
\nabla \cdot [K(E,\vec{r})\nabla f_{e^+}] +
\frac{\partial}{\partial E} [b(E,\vec{r}) f_{e^+}]+Q_{e^+}(E,\vec{r})\;,
\label{transport}
\end{equation}
where convection and annihilations
in the Galactic disk are neglected.
The boundary conditions require the solution 
$f_{e^+}(E,\vec{r},t)$ to vanish at the boundary
of the diffusion zone, which is approximated by a cylinder with 
half-height $L = 1-15~\rm{kpc}$ and radius $ R = 20 ~\rm{kpc}$.

The first term on the right-hand side of the transport equation
is the diffusion
term, which accounts for the propagation of positrons through the
tangled Galactic magnetic fields.
The diffusion coefficient $K(E,\vec{r})$ is assumed to be constant
throughout the diffusion zone and is parametrized by:
\begin{equation}
K(E)=K_0 \;\beta\; {\cal R}^\delta\;,
\end{equation}
where $\beta=v/c$ is the velocity and ${\cal R}$ is the rigidity 
of the particle, 
which is defined as the momentum in GeV per unit charge, 
${\cal R}\equiv p({\rm GeV})/Z$.
The normalization $K_0$ and the spectral index $\delta$
of the diffusion coefficient are related to the properties 
of the interstellar medium and can be determined from 
flux measurements of other cosmic-ray species, mainly from 
the Boron-to-Carbon (B/C) ratio~\cite{Maurin:2001sj}. We list
in Table \ref{tab:param-positron} the diffusion parameters for the
propagation models M2, MED and M1 proposed in ~\cite{Delahaye:2007fr},
which are consistent with the observed B/C ratio.
The second term accounts for energy losses due to 
inverse Compton scattering on starlight and the cosmic microwave 
background, as well as synchrotron radiation and ionization. 
The rate of energy loss, $b(E,\vec{r})$, is assumed to be 
a spatially constant function parametrized by $b(E)=\frac{E^2}{E_0\tau_E}$, 
with $E_0=1\;{\rm GeV}$ and $\tau_E=10^{16}\;{\rm s}$.
Lastly, $Q_{e^+}(E,\vec{r})$ is the source term of positrons
from the decay of a dark matter particle with mass $m_{\rm DM}$
and lifetime $\tau_{\rm DM}$:
\begin{equation}
Q_{e^+}(E,\vec{r})=\frac{\rho_{\rm DM}(\vec{r})}{m_{\rm DM}\tau_{\rm DM}}
\frac{dN_{e^+}(E)}{dE}\;,
\label{source-term}
\end{equation}
where $dN_{e^+}/dE$ is the energy spectrum of positrons
produced in the decay and $\rho_{\rm DM}(\vec{r})$
is the density profile of dark matter in our Galaxy. 
For our numerical analysis, we will adopt the spherically symmetric
Navarro, Frenk and White (NFW) profile~\cite{Navarro:1995iw}:
\begin{equation}
\rho_{\rm DM}(r)=\frac{\rho_0}{(r/r_c)
[1+(r/r_c)]^2}\;,
\label{NFW}
\end{equation}
with $\rho_0=0.26~{\rm GeV}/{\rm cm}^3$ and $r_c=20~{\rm kpc}$.
The normalization is chosen such that the local halo density is
$\rho_{\rm DM}(r_\odot) = 0.3~{\rm GeV}/{\rm cm}^3$ with
$r_\odot = 8.5~{\rm kpc}$~\cite{Bergstrom:1997fj}.

\begin{table}[t]
\begin{center}
\begin{tabular}{|c|ccc|}
\hline
Model & $\delta$ & $K_0\,({\rm kpc}^2/{\rm Myr})$ & $L\,({\rm kpc})$ \\
\hline 
M2 & 0.55 & 0.00595 & 1  \\
MED & 0.70 & 0.0112 & 4 \\
M1 & 0.46 & 0.0765 & 15  \\
\hline
\end{tabular}
\caption{\label{tab:param-positron}\small 
Diffusion parameters for the propagation models M2, MED and M1 
proposed in ~\cite{Delahaye:2007fr}
which are consistent with the observed B/C ratio.
}
\end{center}
\end{table}

The solution of the transport equation at the Solar System, 
$r=r_\odot$, $z=0$, can be formally expressed by the convolution
\begin{equation}
f_{e^+}(E)=\frac{1}{m_{\rm DM} \tau_{\rm DM}}
\int_0^{m_{\rm DM}}dE^\prime G_{e^+}(E,E^\prime) 
 \frac{dN_{e^+}(E^\prime)}{dE^\prime}\;.
\label{solution}
\end{equation}
The solution is thus factorized into two parts.
The first part, given by the Green's function $G(E,E^\prime)$,
encodes all of the information about the astrophysics 
(such as the details of the halo profile and the 
complicated propagation of positrons in the Galaxy) 
and is universal for any decaying dark matter candidate. The
remaining part depends exclusively on the nature and properties
of the decaying dark matter candidate, namely the mass, the lifetime 
and the energy spectrum of positrons produced in the decay.
In the next Section we will analyze several phenomenological scenarios 
of decaying dark matter and discuss their viability in view
of the PAMELA data.

The explicit form of the Green's function is~\cite{Hisano:2005ec}
\begin{equation}
G_{e^+}(E,E^\prime)=\sum_{n,m=1}^\infty 
B_{nm}(E,E^\prime) 
J_0\left(\zeta_n \frac{r_\odot}{R}\right) 
\sin\left(\frac{m \pi}{2 }\right),
\label{greens-function}
\end{equation}
where $J_0$ is the zeroth-order Bessel function of the first kind, whose
successive zeros are denoted by $\zeta_n$. On the other hand,
\begin{equation}
B_{nm}(E,E^\prime)=\frac{\tau_E E_0}{E^2}
C_{nm} 
\exp\left\{\left(\frac{\zeta_n^2}{R^2} + \frac{m^2 \pi^2}{4 L^2}\right) 
\frac{K_0 \tau_E}{\delta - 1} 
\left[\left(\frac{E}{E_0}\right)^{\delta-1}
-\left(\frac{{E^\prime}}{E_0}\right)^{\delta-1}\right]\right\},
\end{equation}
with
\begin{equation}
C_{nm}=\frac{2}{J_1^2(\zeta_n)R^2 L} \int_0^R r^\prime dr^\prime 
\int_{-L}^L dz^\prime  \rho_{\rm DM}(\vec{r}\,^\prime)
J_0\left(\zeta_n \frac{r^\prime}{R}\right) 
\sin\left[\frac{m \pi}{2 L}(L-z^\prime )\right]\;,
\end{equation}
where $J_1$ is the first-order Bessel function.

The Green's function can be well approximated by the following
interpolating function, which is valid for any decaying dark 
matter particle~\cite{Ibarra:2008qg}:
\begin{equation}
G_{e^+}(E,E^\prime)\simeq\frac{10^{16}}{E^2}
e^{a+b(E^{\delta-1}-E^{\prime \delta-1})}
\theta(E^\prime-E)\,{\rm cm}^{-3}\,{\rm s}\;,
\label{interp-pos}
\end{equation}
where $E$ and $E^\prime$ are expressed in units of GeV.
The coefficients $a$ and $b$ can be found in
Table~\ref{tab:fit-positron} for the NFW profile and 
the different diffusion models listed in Table~\ref{tab:param-positron}.
This approximation works better than 15 -- 20\% over the whole range
of energies.
We find numerically that the Green's function is not very sensitive
to the choice of the halo profile, since
the Earth receives only positrons created within a few kpc 
from the Sun, where the different halo profiles are very similar.
Therefore, the Green's function coefficients
for other halo profiles can be well approximated by the values
in Table~\ref{tab:fit-positron}.

\begin{table}[t]
\begin{center}
\begin{tabular}{|c|cc|}
 \hline
model & $a$ & $b$ \\ 
\hline
M2 & $-0.9716$ & $-10.012$  \\
MED & $-1.0203$ & $-1.4493$  \\
M1 & $-0.9809$ & $-1.1456$  \\
 \hline
\end{tabular}
\caption{\label{tab:fit-positron}\small 
Coefficients of the interpolating function Eq.~(\ref{interp-pos}) 
for the positron Green's function, assuming a NFW halo profile
and for the different diffusion models in Table~\ref{tab:param-positron}.}
\end{center}
\end{table}

Finally, the flux of primary positrons at the Solar System
from dark matter decay is given by:
\begin{equation}
\Phi_{e^+}^{\rm{prim}}(E) = \frac{c}{4 \pi} f_{e^+}(E).
\label{flux}
\end{equation}

In order to compare our results with experiments, we will calculate
the positron fraction, which is defined as the ratio of the total positron
flux over the total electron plus positron fluxes, 
$\Phi_{e^+}/(\Phi_{e^-}+\Phi_{e^+})$.
The total positron flux receives contributions from 
the dark matter decay as well as from a secondary component
stemming from the collision of primary
protons and other nuclei on the interstellar medium, 
which constitutes the background to any dark matter signal. 
On the other hand, the total electron flux has a primary
astrophysical component, presumably originating from supernova
remnants, a secondary component from spallation of cosmic rays 
on the interstellar medium,
and an exotic primary component from dark matter decay that
might be important at high energies in some scenarios.
For the background fluxes of primary and secondary electrons, 
as well as secondary positrons, we will use the parametrizations 
obtained in \cite{Baltz:1998xv} from detailed computer 
simulations of cosmic-ray propagation~\cite{Moskalenko:1997gh}:
\begin{eqnarray}
\Phi_{e^-}^{\rm{prim}}(E) &=& \frac{0.16 \,E^{-1.1}}
{1 + 11 \,E^{0.9} + 3.2 \,E^{2.15}} 
~(\rm{GeV}^{-1} \rm{cm}^{-2} \rm{s}^{-1} \rm{sr}^{-1})\;,\\
\Phi_{e^-}^{\rm{sec}}(E) &=& \frac{0.70 \,E^{0.7}}
{1 + 110 \,E^{1.5} + 600 \,E^{2.9} + 580 \,E^{4.2}} 
~(\rm{GeV}^{-1} \rm{cm}^{-2} \rm{s}^{-1} \rm{sr}^{-1})\;,\\
\Phi_{e^+}^{\rm{sec}}(E) &=& \frac{4.5 \,E^{0.7}}
{1 + 650 \,E^{2.3} + 1500 \,E^{4.2}} 
~(\rm{GeV}^{-1} \rm{cm}^{-2} \rm{s}^{-1} \rm{sr}^{-1})\;,
\end{eqnarray}
where $E$ is the energy expressed in units of GeV.

\section{Results}

For a given model of decaying dark matter and a given propagation model,
defined by the parameters in Table \ref{tab:param-positron}, the 
predictions for the positron fraction at Earth can be readily
computed using the formalism explained in the previous section.
To derive our results we solved the transport equation 
Eq.~(\ref{transport}) numerically, although for a quick estimate
of the positron fraction the interpolating function Eq.~(\ref{interp-pos})
may also be used.

To keep the analysis as model-independent as possible, we will
analyze several possibilities for the decaying dark matter,
computing the prediction for the positron fraction for
either a fermionic or a bosonic particle which decays
in various channels with a branching ratio of 100\%. 
We will treat the dark matter mass and lifetime
as free parameters, while the normalization of the background is kept
fixed.
We will first adopt the MED propagation
model in order to better compare the predictions 
from different particle physics scenarios and later on, 
we will analyze the sensitivity of the predictions to the 
choice of the propagation model.

Let us now discuss the cases of fermionic and scalar
dark matter particles separately.\\

In the case that the dark matter particle is a fermion $\psi$,
the following decay channels are possible:
\begin{eqnarray}
\psi&\rightarrow& Z^0 \nu\;, \nonumber \\ 
\psi&\rightarrow& W^\pm \ell^\mp\;, \nonumber\\
\psi&\rightarrow& \ell^+ \ell^- \nu\;,
\end{eqnarray}
provided that the respective channels are kinematically open. 

\begin{figure}[t]
\begin{center}
\begin{tabular}{c}
\psfig{figure=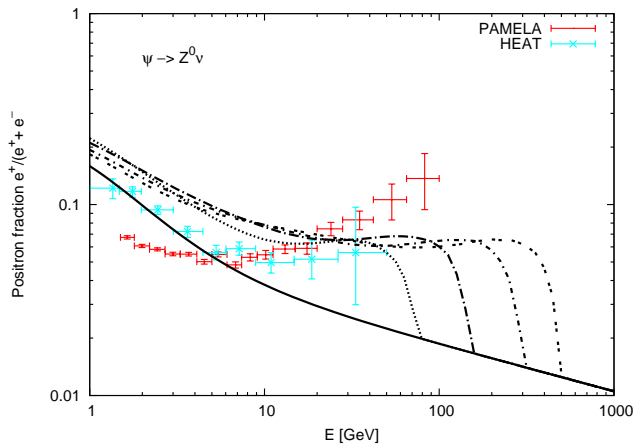,width=85mm} 
\end{tabular}
\end{center}
\caption{\label{ferm-Znu}\small 
Positron fraction from the decay of the fermionic dark
matter particle in the channel $\psi\rightarrow Z^0 \nu$ when 
the dark matter mass is, from left to right, 
$m_{\rm DM}= 150,\;300,\; 600,\; 1000\;{\rm GeV}$.
The lifetime has been chosen to provide a qualitatively good fit
to the data and is $\sim 5\times 10^{25}\;{\rm s}$ in all the
cases.}
\end{figure}

The fragmentation of the $Z^0$ boson in the 
decay $\psi\rightarrow Z^0 \nu$ produces
a continuous spectrum of positrons (mainly from
$\pi^+$ decay) that we have obtained using the 
event generator PYTHIA 6.4~\cite{Sjostrand:2006za}\footnote{
The fragmentation of the weak gauge bosons 
also produces fluxes of primary antiprotons and gamma-rays,
which are severely constrained by the PAMELA~\cite{Adriani:2008zq}
and EGRET~\cite{egret} experiments, respectively.
The predictions for the antiproton and the gamma-ray
fluxes of the phenomenological scenarios studied in this paper 
will be presented elsewhere~\cite{inpreparation}.}. 
The predicted positron fraction is shown in Fig.~\ref{ferm-Znu}, compared
to the PAMELA and HEAT data, for the MED propagation model
and for different dark matter masses. The lifetime has been
chosen in order to produce a qualitatively good fit of
the prediction to the data points. 
We operate under the assumption that 
the difference between the low-energy data from HEAT and PAMELA 
is due to solar modulation. Therefore, we use the HEAT data at low 
energies, which were recorded during a period of minimum solar activity.
It is apparent from the figure that this decay channel
by itself cannot explain the steep rise of the spectrum
observed by PAMELA for any of the masses analyzed here. 
We have also checked that using different propagation
models does not improve the fit to the data significantly.

\begin{figure}[t]
\begin{center}
\begin{tabular}{c}
\hspace{-1.0cm}
\psfig{figure=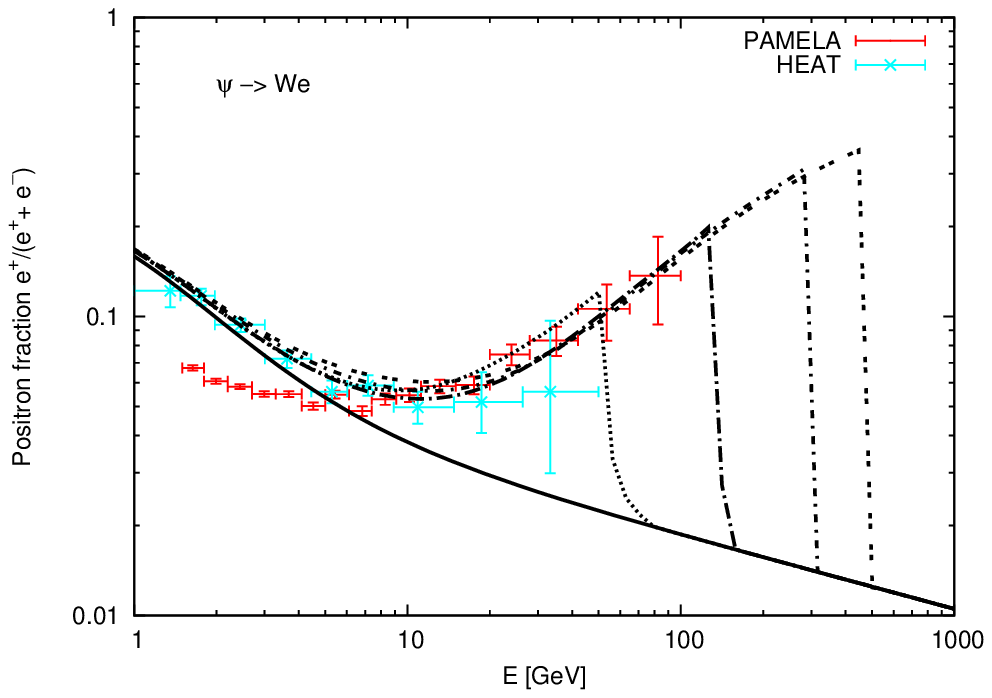,width=85mm} 
\psfig{figure=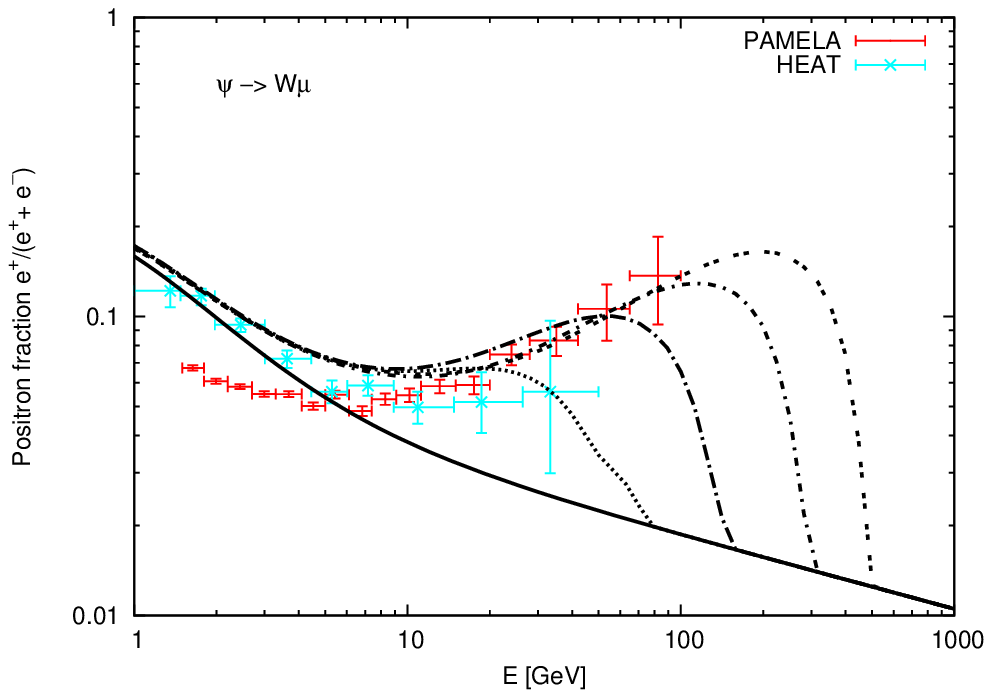,width=85mm} \\
\hspace{-1.0cm}
\psfig{figure=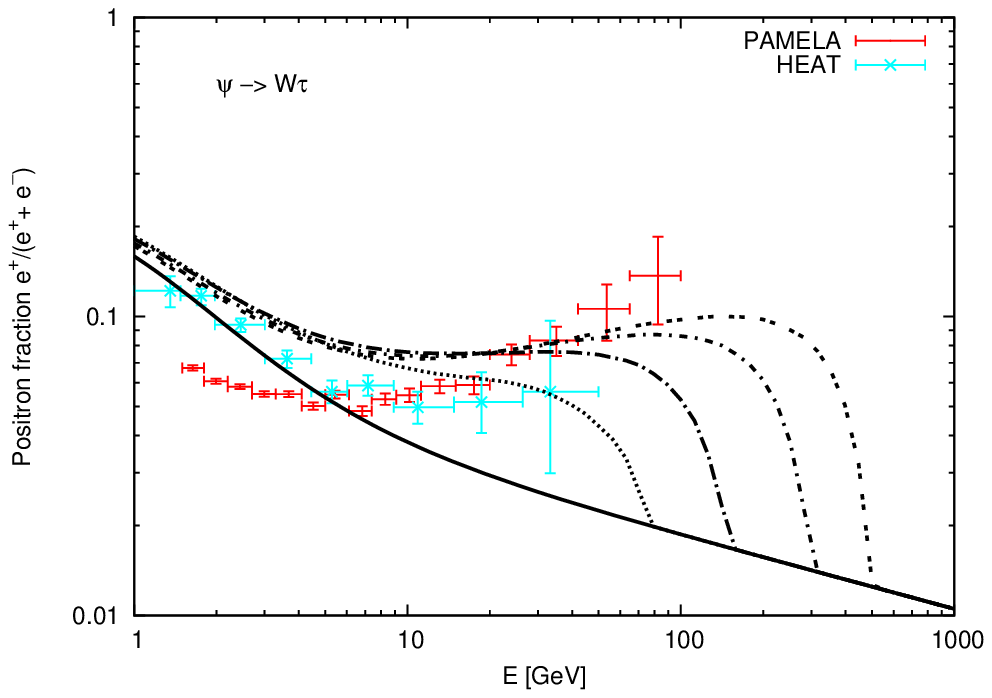,width=85mm} 
\end{tabular}
\end{center}
\caption{\label{ferm-Wl}\small 
Positron fraction from the decay of the fermionic dark
matter particle in the channels $\psi\rightarrow W^\pm e^\mp$ (top-left panel),
$\psi\rightarrow W^\pm \mu^\mp$ (top-right panel)
and $\psi\rightarrow W^\pm \tau^\mp$ (bottom panel),
when the dark matter mass is, from left to right, 
$m_{\rm DM}= 150,\;300,\; 600,\; 1000\;{\rm GeV}$.
The lifetime, which ranges between $10^{26}\;{\rm s}$ 
and $5\times 10^{26}\;{\rm s}$, is different in each case and has been 
chosen to provide a qualitatively good fit to the data.
}
\end{figure}

On the other hand, we show in Fig.~\ref{ferm-Wl} the prediction
for the positron fraction when the fermionic dark
matter particle decays as $\psi\rightarrow W \ell$.
The positrons created in the fragmentation of the 
$W$ gauge bosons produce a rather flat contribution
to the positron fraction.
However, the hard positrons resulting
from the decay of the $\mu$ and $\tau$ leptons or directly 
from the dark matter decay into positrons produce a rise in
the spectrum, which is most prominent in the decay mode 
$\psi\rightarrow W^\pm e^\mp$, although it is also quite visible
in $\psi\rightarrow W^\pm \mu^\mp$.
Interestingly, these two decay models can qualitatively 
reproduce the energy spectrum measured by PAMELA
for dark matter masses larger than 
$\sim 300\;{\rm GeV}$. 
Future measurements of the positron fraction at energies
100 -- 300 GeV, as planned by the PAMELA collaboration,
will be crucial to discriminate among these  
possibilities. Note that when the dark matter
mass is large, the positron fraction in the scenario
with direct decay into positrons, $\psi\rightarrow W^\pm e^\mp$, can reach values
as large as 30 -- 40\%. In this case, not only the positron flux, 
but also the {\it electron} flux will receive a significant
primary contribution from dark matter decay.

\begin{figure}[t]
\begin{center}
\begin{tabular}{c}
\hspace{-1.0cm}
\psfig{figure=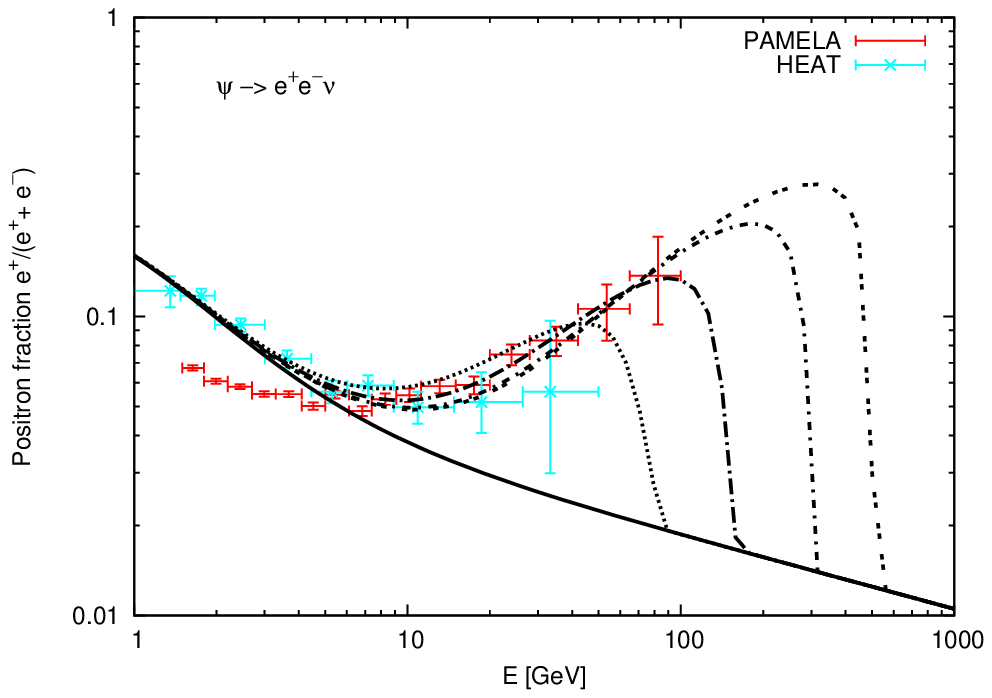,width=85mm} 
\psfig{figure=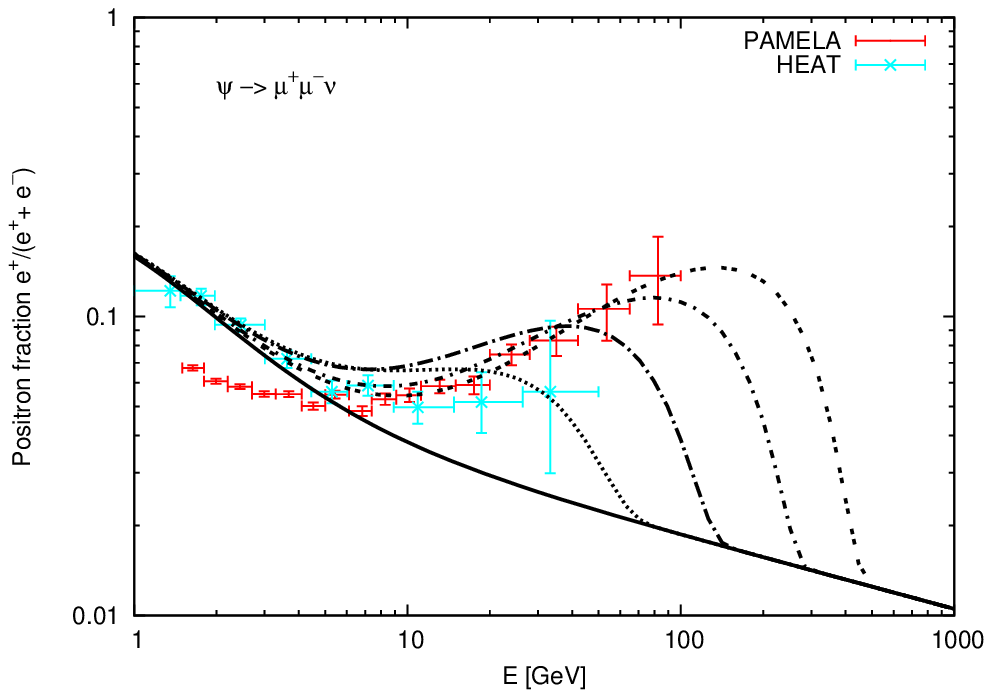,width=85mm} \\
\hspace{-1.0cm}
\psfig{figure=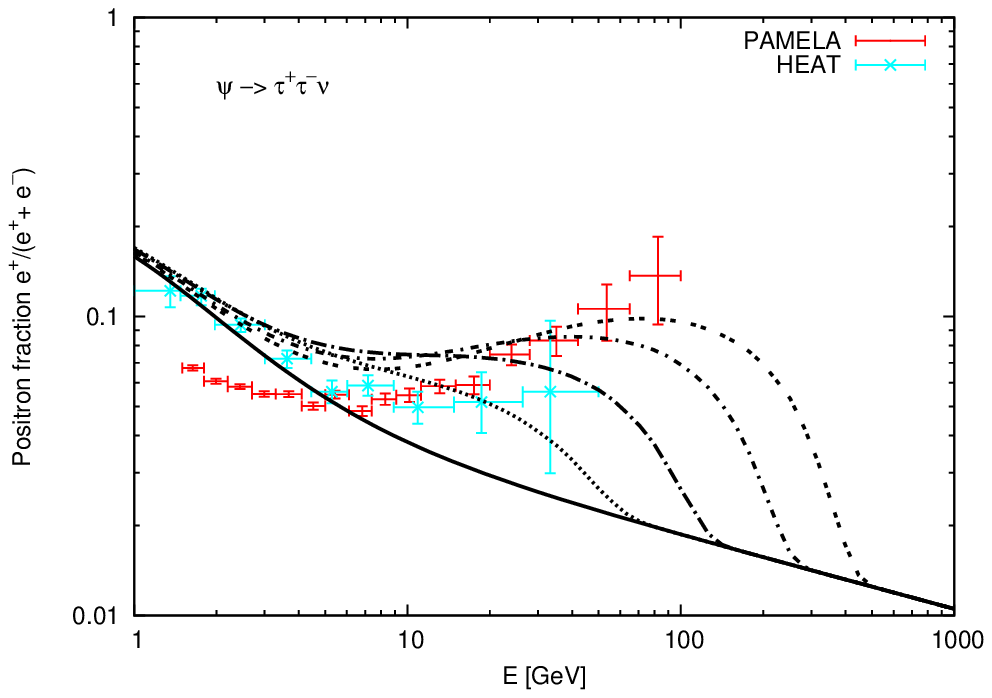,width=85mm} 
\end{tabular}
\end{center}\caption{\label{ferm-3body}\small 
Positron fraction from the decay of the fermionic dark
matter particle in the channels $\psi\rightarrow e^+ e^- \nu$ (top-left panel),
$\psi\rightarrow  \mu^+ \mu^- \nu$ (top-right panel)
and $\psi\rightarrow \tau^+ \tau^- \nu$ (bottom panel),
when the dark matter mass is, from left to right, 
$m_{\rm DM}= 150,\;300,\; 600,\; 1000\;{\rm GeV}$.
The lifetime, which ranges between $5\times 10^{25}\;{\rm s}$ 
and $8\times 10^{26}\;{\rm s}$, is different in each case and has been 
chosen to provide a qualitatively good fit to the data.}
\end{figure}

Lastly, we have also analyzed the case that the dark matter
particle decays leptonically in a three-body decay,
$\psi \rightarrow \ell^+ \ell^- \nu$;
the results are shown in Fig.~\ref{ferm-3body}.
The decay $\psi \rightarrow e^+ e^- \nu$ can 
explain the PAMELA anomaly when the dark matter
mass is larger than $\sim 300\;{\rm GeV}$,
while the decay channel  $\psi \rightarrow \mu^+ \mu^- \nu$
requires larger masses. In contrast, the energy spectrum
produced in the decay  $\psi \rightarrow \tau^+ \tau^- \nu$,
although it exhibits a notable bump, 
is too flat to explain the observations by itself. In this
case, other contributions to the positron flux should be invoked, 
for instance from pulsars, in order to reproduce the observations.\\

When the decaying dark matter particle is a scalar,
the following decay modes are possible:
\begin{eqnarray}
\phi&\rightarrow& Z^0 Z^0 \nonumber \\ 
\phi&\rightarrow& W^+ W^-,  \nonumber\\
\phi&\rightarrow& \ell^+ \ell^- ,
\end{eqnarray}
again provided that the decays are kinematically open.

\begin{figure}[t]
\begin{center}
\begin{tabular}{c}
\hspace{-1.0cm}
\psfig{figure=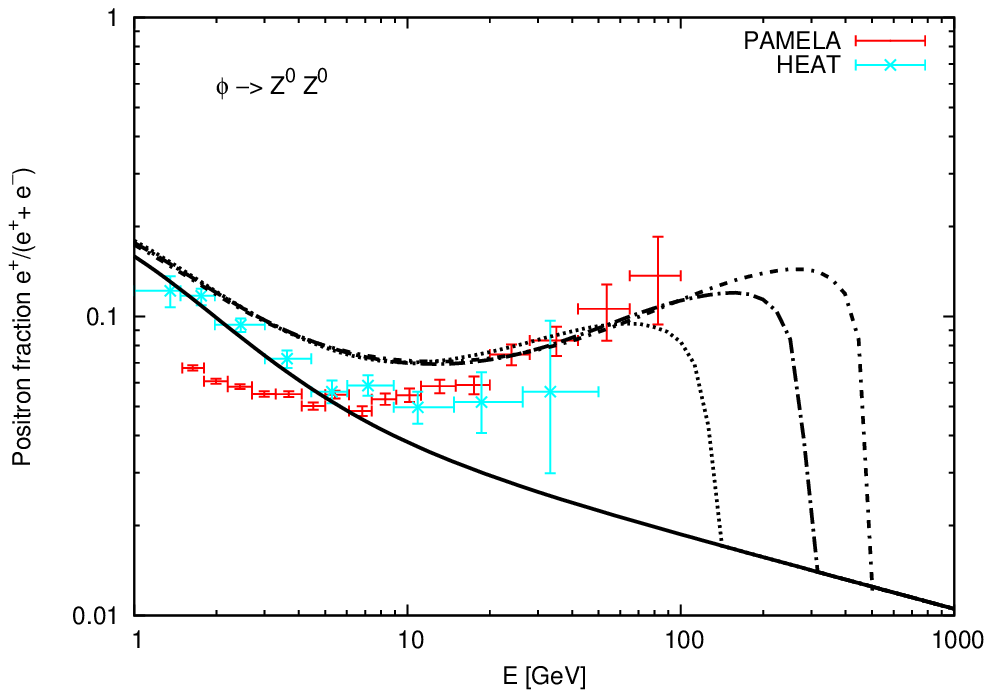,width=85mm} 
\psfig{figure=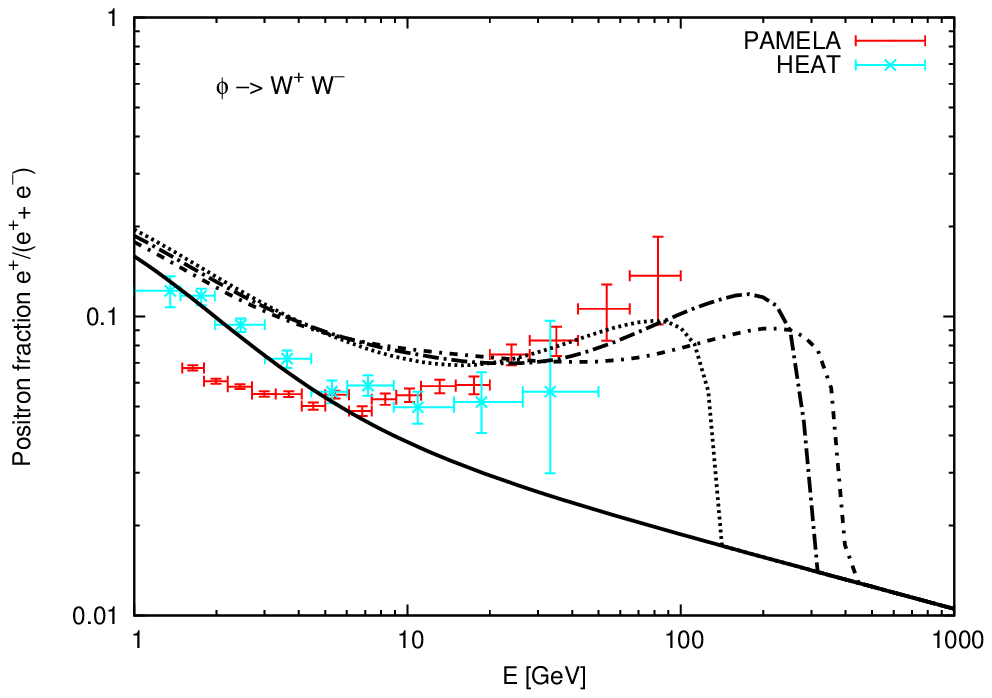,width=85mm} 
\end{tabular}
\end{center}\caption{\label{scalar-gauge}\small 
Positron fraction from the decay of a scalar dark
matter particle in the channels $\phi\rightarrow  Z^0 Z^0$ 
(top-left panel) and $\phi\rightarrow W^+ W^-$ (top-right panel)
when the dark matter mass is, from left to right, 
$m_{\rm DM}= 300,\; 600,\; 1000\;{\rm GeV}$.
The lifetime has been chosen to provide a qualitatively good fit
to the data and is $\sim 2\times 10^{26}\;{\rm s}$ 
($\sim 10^{26}\;{\rm s}$) for the decay
into $Z$ ($W$) bosons.}
\end{figure}

We show in Fig.~\ref{scalar-gauge} the expected positron fraction from 
the decay of a scalar particle into weak gauge bosons,
$\phi \rightarrow Z^0 Z^0$ or $\phi\rightarrow W^+ W^-$.
In these decay modes, no hard lepton is produced. Instead,
only positrons from the fragmentation of the weak gauge
bosons will contribute. As a result, the spectral shape of the positron
fraction is too flat to explain the steep rise
observed by PAMELA by itself.

\begin{figure}[t]
\begin{center}
\begin{tabular}{c}
\hspace{-1.0cm}
\psfig{figure=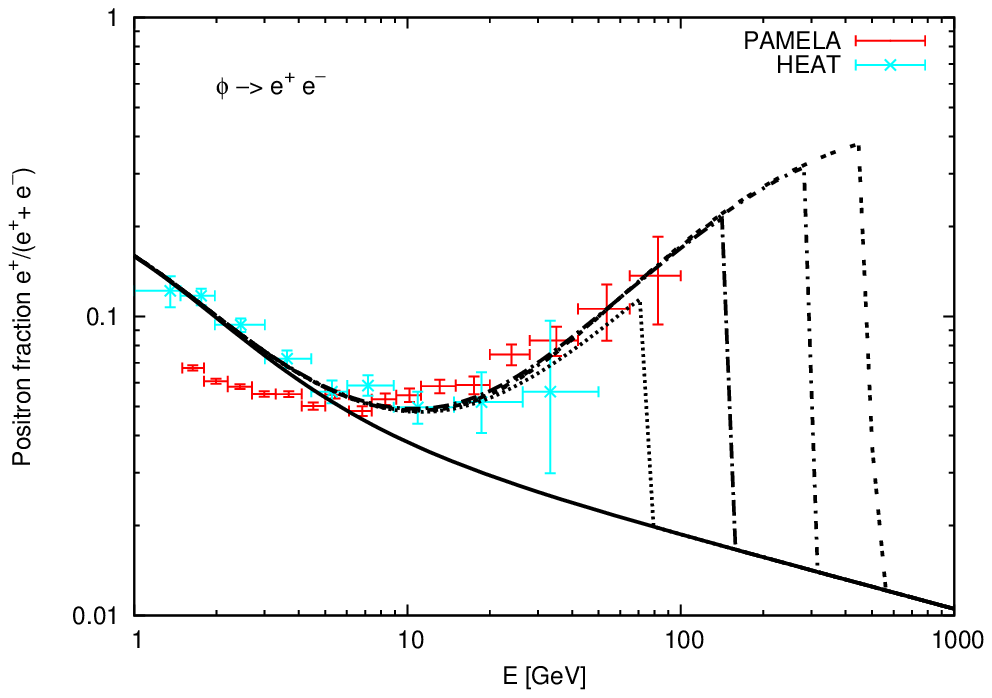,width=85mm} 
\psfig{figure=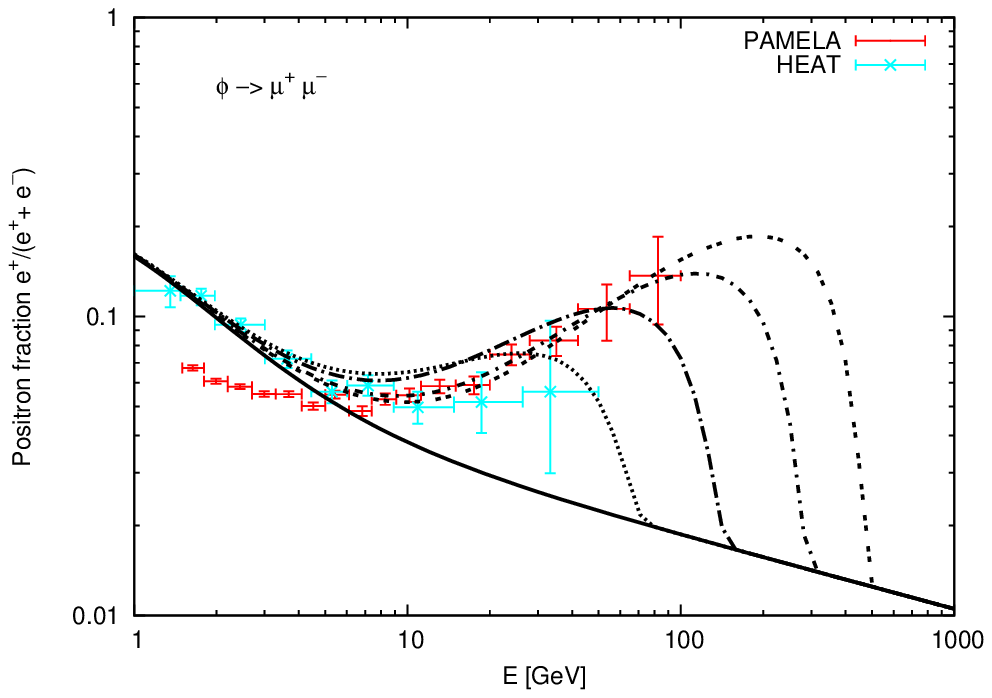,width=85mm} \\
\hspace{-1.0cm}
\psfig{figure=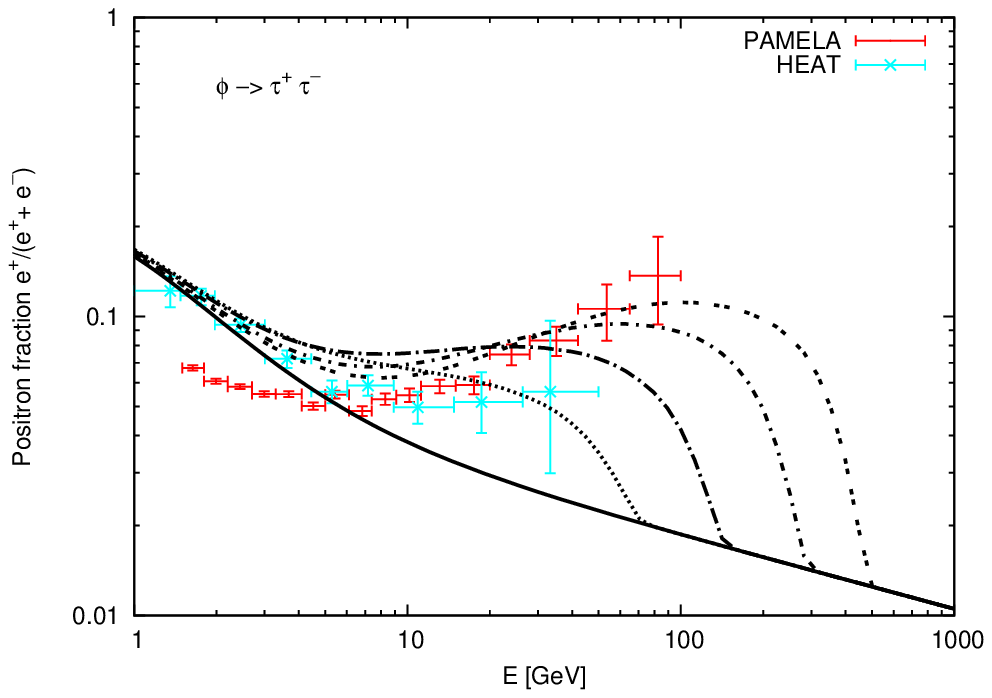,width=85mm} 
\end{tabular}
\end{center}\caption{\label{scalar-lepton}\small 
Positron fraction from the decay of a scalar dark
matter particle in the channels 
$\phi\rightarrow e^+ e^-$ (top-left panel),
$\phi\rightarrow  \mu^+ \mu^-$ (top-right panel)
and $\phi\rightarrow \tau^+ \tau^-$ (bottom panel),
when the dark matter mass is, from left to right, 
$m_{\rm DM}= 150,\;300,\; 600,\; 1000\;{\rm GeV}$.
The lifetime, which ranges between $10^{26}\;{\rm s}$
and $10^{27}\;{\rm s}$, is different in each case and has been 
chosen to provide a qualitatively good fit to the data.}
\end{figure}

We have also calculated the predictions for the positron
fraction when the scalar dark matter decays directly into two
charged leptons, $\phi \rightarrow \ell^+ \ell^-$; the
results are shown in Fig.~\ref{scalar-lepton}.
The predictions are qualitatively very similar to the 
ones in the scenario where a dark matter
fermion decays into $W^\pm \ell^\mp$, the most noticeable 
difference being the larger cutoff
in the energy spectrum in the scalar case, for the same dark matter mass. 
To be more precise, in the scenario with 
$\psi\rightarrow W^\pm \ell^\mp$ the cutoff is
at $m_{\rm DM}/2(1-M^2_W/m^2_{\rm DM})$ while in
the scenario with $\phi\rightarrow \ell^+ \ell^-$,
it is at the larger value $m_{\rm DM}/2$.
For this class of scenarios, the decay modes into
positrons or antimuons can accommodate the PAMELA
anomaly better than the decay mode into antitaus.\\

Above we have analyzed the
predictions for the positron fraction for 
different scenarios of fermionic or scalar decaying
dark matter, assuming the MED propagation model.
After discussing the sensitivity of the predictions
to the choice of the particle physics model, we would like to 
briefly address the sensitivity of the results to astrophysical uncertainties.

\begin{figure}[t]
\begin{center}
\begin{tabular}{c}
\hspace{-1.0cm}
\psfig{figure=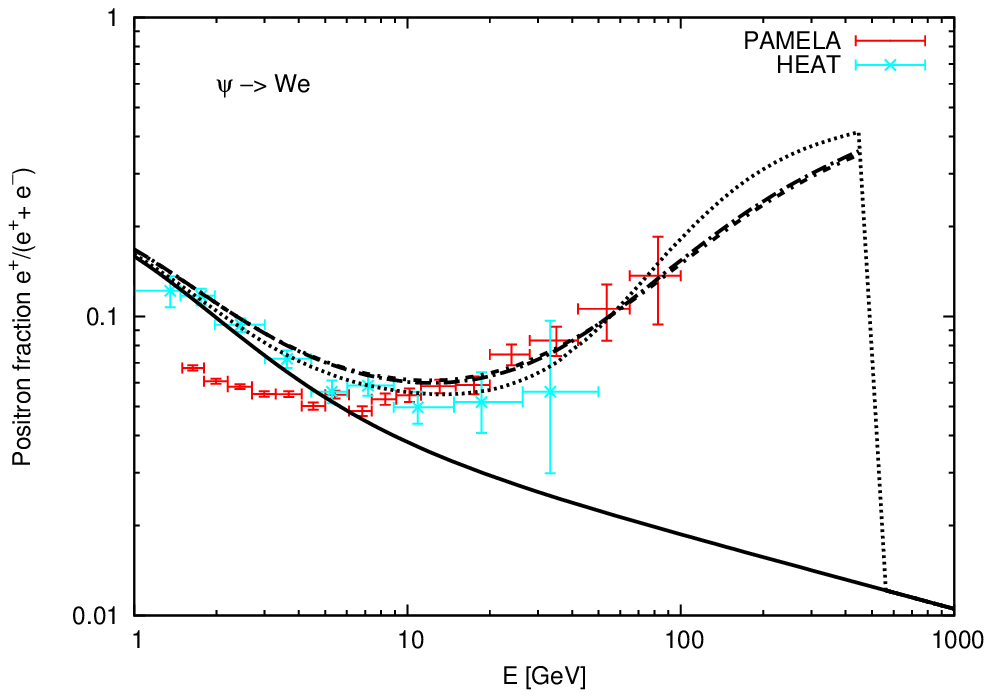,width=85mm} 
\psfig{figure=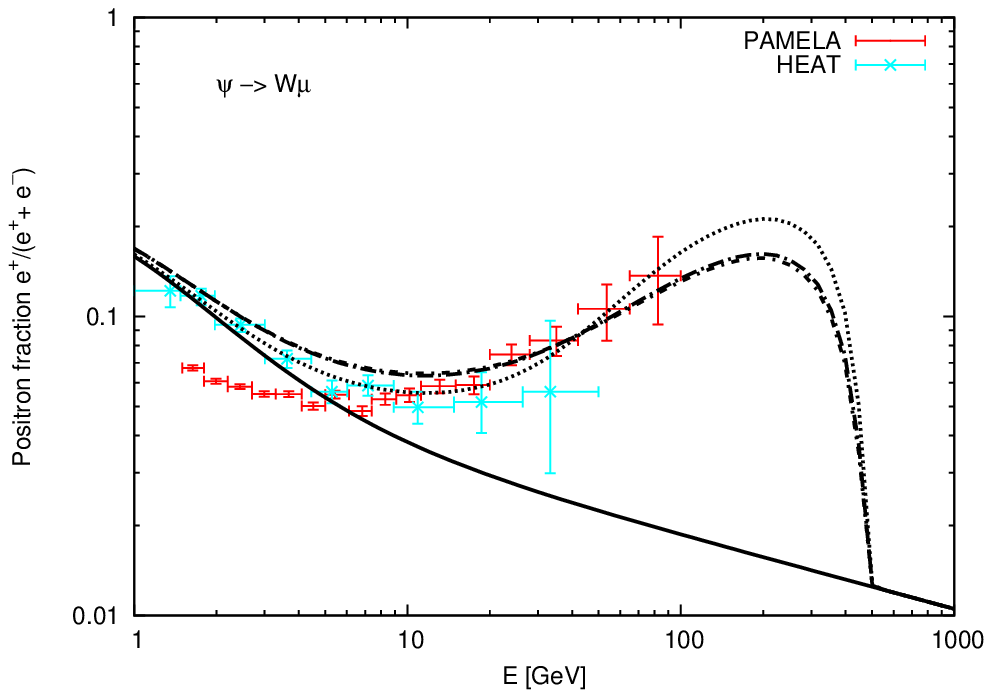,width=85mm} \\
\hspace{-1.0cm}
\psfig{figure=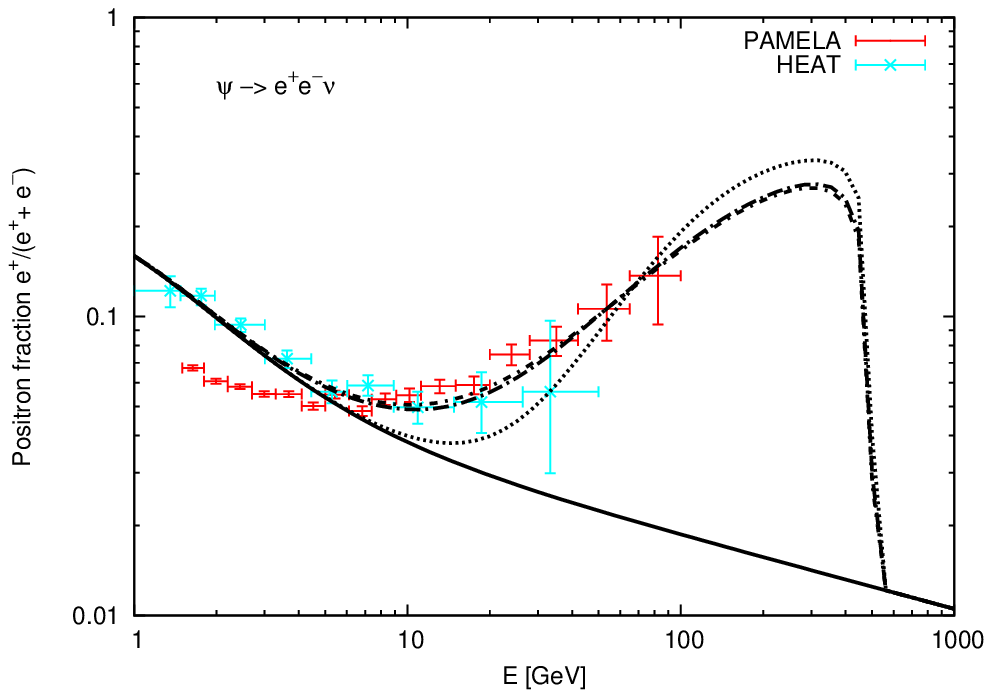,width=85mm} 
\psfig{figure=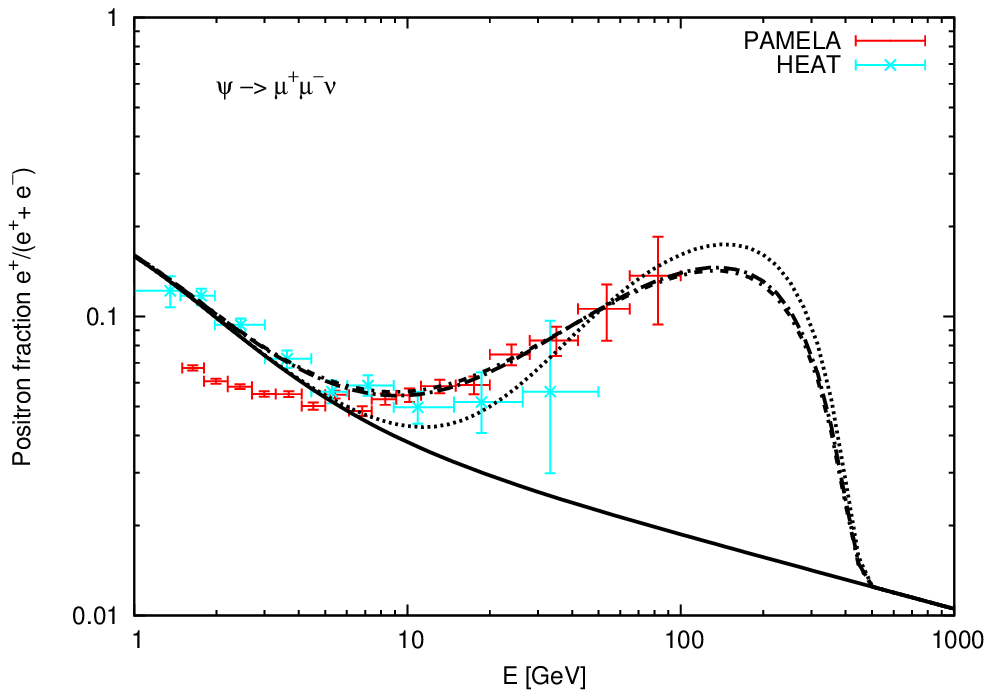,width=85mm} \end{tabular}
\end{center}\caption{\label{models}\small 
Dependence of the predicted positron fraction on the
propagation model. For a decaying dark matter fermion
with $m_{\rm DM}= 1000\;{\rm GeV}$, we show the
decay modes 
$\psi\rightarrow W^\pm e^\mp $ (top-left panel),
$\psi\rightarrow W^\pm \mu^\mp$ (top-right panel)
$\psi\rightarrow e^+ e^- \nu$ (bottom-left panel),
$\psi\rightarrow  \mu^+ \mu^- \nu$ (bottom-right panel).
The dotted line corresponds to the M2 propagation model,
while the almost indistinguishable dashed-dotted lines correspond to the MED and the M1 models
(see Table \ref{tab:param-positron}). }
\end{figure}

We show in Fig.~\ref{models} the sensitivity
to the propagation model of four particle physics 
scenarios consisting of a dark matter fermion which
decays into two particles, $\psi\rightarrow W^\pm (e,\mu)^\mp $, 
or into three particles, $\psi\rightarrow (e,\mu)^+ (e,\mu)^- \nu$.
These scenarios are characterized by the direct decay
into hard  positrons or antimuons, and were found 
to be compatible with the excess observed by PAMELA 
for the MED propagation model. 
The results in Fig.~\ref{models} show that this
conclusion still holds for the M1 and M2 propagation models:
the M1 and MED propagation models yield very similar
predictions of the positron fraction, while the M2 propagation
model yields a slightly larger positron fraction at high energies
and a smaller positron fraction at low energies. The reason
is that in the M2 propagation model the diffusion is
minimal and thus the hard spectrum of injected positrons
from dark matter decay is less altered by the propagation than
in the MED and M1 model, yielding a steeper rise in the
positron fraction measured at Earth. 
The same conclusion holds for the scenarios where a
scalar dark matter particle directly decays into an
electron-positron pair or a muon-antimuon pair,
$\phi\rightarrow (e,\mu)^+ (e,\mu)^-$,
which as discussed above yield very similar signatures in
the positron fraction as the scenarios with fermionic 
dark matter $\psi\rightarrow W^\pm (e,\mu)^\mp$,
where the dark matter also decays into a monoenergetic 
positron or antimuon.\\

As a last remark, let us note that the 
origin of the positron excess
might not be the decay of the dark matter itself, but the decay
of another long-lived particle which is present in our Galaxy. 
Notice that in the scenarios
analyzed in this paper with decaying dark matter, the
shape of the spectrum is 
determined by the dark matter mass, while the normalization, by
the combination $\rho_{\rm DM}(r_\odot)/\tau_{\rm DM}$, with
$\rho_{\rm DM}(r_\odot) = 0.3~{\rm GeV}/{\rm cm}^3$ being the
local density of dark matter particles
({\it cf.} Eqs.~(\ref{source-term},\ref{NFW})). 
Therefore, identical signatures in the positron fraction are
obtained if the decaying particle has a mass  $m^\prime$,
a lifetime $\tau^\prime$ and a local abundance $\rho^\prime(r_\odot)$
satisfying\footnote{We assume here that any 
weakly or superweakly interacting particle with a lifetime
longer than the age of the Universe is present in our Galaxy
and has a halo distribution identical to the 
dominant component of dark matter.}
\begin{eqnarray}
m^\prime&=&m_{\rm DM} \;, \nonumber \\
\rho^\prime(r_\odot)/\tau^\prime&=&\rho_{\rm DM}(r_\odot)/\tau_{\rm DM}\;.
\end{eqnarray}
This opens new possibilities for model building. For instance,
it is conceivable that the dominant component of dark matter
is completely stable (perhaps even annihilating), while the origin
of the positron excess is the decay of a subdominant
form of dark matter.

\section{Conclusions}

We have computed the predictions for
the positron fraction of several decaying 
dark matter scenarios and discussed their viability
in view of the new measurements reported by
the PAMELA collaboration.
We have studied scenarios where the dark matter particle
is a fermion $\psi$, which decays in the two-body decay channels
$\psi\rightarrow Z^0 \nu,\;W^\pm \ell^\mp$ or in 
the three-body decay channel $\psi\rightarrow \ell^+ \ell^- \nu$,  
where $\ell=e,\;\mu,\;\tau$ denotes the charged leptons. In addition,
we have also studied scenarios where the dark matter particle
is a scalar $\phi$, which decays as
$\phi \rightarrow Z^0 Z^0,\; \phi \rightarrow W^+ W^-,\; \phi \rightarrow \ell^+ \ell^-$.

We have found that decay channels producing hard positrons
and antimuons are favored by the PAMELA data. 
Namely, the decay channels
of a fermionic dark matter particle $\psi\rightarrow W (\mu,e)$
and $\psi\rightarrow (\mu,e)^+ (\mu,e)^- \nu$, and the decay
channels of a scalar dark matter particle 
$\phi\rightarrow (\mu,e)^+ (\mu,e)^-$ produce a steep
rise in the positron fraction above 10 GeV, as 
observed by the PAMELA collaboration. It is worth noting that 
the spectral shape of the positron fraction resulting from 
a sharply peaked positron or antimuon injection spectrum 
matches the experimental results quite accurately.
On the other
hand, decay modes involving the tau lepton can contribute to 
an excess in the positron fraction at high energies,
although the resulting spectrum is too flat. Finally,
the decay of a scalar particle into weak gauge bosons,
$\phi \rightarrow Z^0 Z^0,\; \phi \rightarrow W^+ W^-$, can also
contribute to an excess in the positron fraction, although 
the spectrum is also too flat to fit the PAMELA data by itself.
In addition, these decay channels are also disfavored by antiproton
overproduction constraints for dark matter lifetimes that yield 
appreciable contributions to the positron flux.

If dark matter decay is indeed the primary cause for the observed 
positron excess, the PAMELA data
clearly seem to point toward a rather heavy dark matter particle, 
$m_{\text{DM}} \gsim 300~\text{GeV}$, which preferentially decays directly into 
first or second generation charged leptons with a lifetime 
$\tau_{\text{DM}} \sim 10^{26}~\text{s}$.

Future measurements of the positron fraction in the energy
range 100 -- 300 GeV by PAMELA will provide 
invaluable information 
about scenarios with decaying dark matter and the properties
of the decaying particles. Furthermore, 
the constraints on the predicted antiproton flux
by these same scenarios from PAMELA~\cite{Adriani:2008zq}
and on the gamma-ray flux from EGRET~\cite{egret} (and in 
the near future from the Fermi Gamma-ray Space 
Telescope~\cite{FGST}) offer complementary information
about this scenario, which will be presented elsewhere~\cite{inpreparation}.

\section*{Note Added}

During the completion of this work two preprints appeared
presenting related analyses \cite{Yin:2008bs,Ishiwata:2008cv}.


\begin{thebibliography}{99}

\bibitem{Moskalenko:1997gh}
  I.~V.~Moskalenko and A.~W.~Strong,
  Astrophys.\ J.\  {\bf 493} (1998) 694.
%
%
%
\bibitem{Barwick:1997ig}
  S.~W.~Barwick {\it et al.}  [HEAT Collaboration],
  Astrophys.\ J.\  {\bf 482} (1997) L191.
%
%
%
\bibitem{CAPRICE}
  M.~Boezio {\it et al.}  [CAPRICE Collaboration],
  Astrophys.\ J.\  {\bf 532} (2000) 653.
%
%
%
\bibitem{Grimani:2002yz}
  C.~Grimani {\it et al.},
  Astron.\ Astrophys.\  {\bf 392} (2002) 287.
%
%
%
\bibitem{Aguilar:2007yf}
  M.~Aguilar {\it et al.}  [AMS-01 Collaboration],
  Phys.\ Lett.\  B {\bf 646}, 145 (2007).
%
%
%
\bibitem{annihilating}
  A.~J.~Tylka,
  Phys.\ Rev.\ Lett.\  {\bf 63} (1989) 840
  [Erratum-ibid.\  {\bf 63} (1989) 1658];
%
  M.~S.~Turner and F.~Wilczek,
  Phys.\ Rev.\  D {\bf 42} (1990) 1001;
%
  M.~Kamionkowski and M.~S.~Turner,
  Phys.\ Rev.\  D {\bf 43} (1991) 1774;
%
  G.~L.~Kane, L.~T.~Wang and J.~D.~Wells,
  Phys.\ Rev.\  D {\bf 65} (2002) 057701;
%
  E.~A.~Baltz, J.~Edsjo, K.~Freese and P.~Gondolo,
  Phys.\ Rev.\  D {\bf 65} (2002) 063511;
%
  G.~L.~Kane, L.~T.~Wang and T.~T.~Wang,
  Phys.\ Lett.\  B {\bf 536} (2002) 263;
%
  H.~C.~Cheng, J.~L.~Feng and K.~T.~Matchev,
  Phys.\ Rev.\ Lett.\  {\bf 89}, 211301 (2002);
%
  D.~Hooper and G.~D.~Kribs,
  Phys.\ Rev.\  D {\bf 70} (2004) 115004;
%
  M.~Cirelli, R.~Franceschini and A.~Strumia,
  Nucl.\ Phys.\  B {\bf 800} (2008) 204.
%
%
%
\bibitem{Baltz:1998xv}
  E.~A.~Baltz and J.~Edsjo,
  Phys.\ Rev.\  D {\bf 59} (1999) 023511.
%
%
%
\bibitem{Hisano:2005ec}
  J.~Hisano, S.~Matsumoto, O.~Saito and M.~Senami,
  Phys.\ Rev.\  D {\bf 73} (2006) 055004.
%
%
%
\bibitem{Buchmuller:2007ui}
  W.~Buchm\"uller, L.~Covi, K.~Hamaguchi, A.~Ibarra and T.~Yanagida,
  JHEP {\bf 0703}, 037 (2007).
%
%
%
\bibitem{Ibarra:2007wg}
  A.~Ibarra and D.~Tran,
  Phys.\ Rev.\ Lett.\  {\bf 100}, 061301 (2008).
%
%
%
\bibitem{Ishiwata:2008cu}
  K.~Ishiwata, S.~Matsumoto and T.~Moroi,
  arXiv:0805.1133 [hep-ph].
%
%
%
\bibitem{Ibarra:2008qg}
  A.~Ibarra and D.~Tran,
  JCAP {\bf 0807} (2008) 002.
%
%
\bibitem{Covi:2008jy}
  L.~Covi, M.~Grefe, A.~Ibarra and D.~Tran,
  arXiv:0809.5030 [hep-ph].
%
%
%
\bibitem{smr05}
  A.~W.~Strong, I.~V.~Moskalenko and O.~Reimer,
  Astrophys.\ J.\  {\bf 613} (2004) 962;
  Astrophys.\ J.\  {\bf 613} (2004) 956.
%
%
%
\bibitem{CTY}
  C.~R.~Chen, F.~Takahashi and T.~T.~Yanagida,
  arXiv:0809.0792 [hep-ph];
  arXiv:0811.0477 [hep-ph].
%
%
%
\bibitem{Ibarra:2008kn}
  A.~Ibarra, A.~Ringwald and C.~Weniger,
  arXiv:0809.3196 [hep-ph].
%
%
%
\bibitem{Chen:2008dh}
  C.~R.~Chen and F.~Takahashi,
  arXiv:0810.4110 [hep-ph].
%
%
%
\bibitem{Hamaguchi:2008rv}
  K.~Hamaguchi, E.~Nakamura, S.~Shirai and T.~T.~Yanagida,
  arXiv:0811.0737 [hep-ph].
%
%
%
\bibitem{Picozza:2006nm}
  P.~Picozza {\it et al.},
  Astropart.\ Phys.\  {\bf 27} (2007) 296.
%
%
%
\bibitem{Adriani:2008zr}
  O.~Adriani {\it et al.},
  arXiv:0810.4995 [astro-ph].
%
%
%
\bibitem{Delahaye:2008ua}
  T.~Delahaye, F.~Donato, N.~Fornengo, J.~Lavalle, R.~Lineros, P.~Salati and R.~Taillet,
  arXiv:0809.5268 [astro-ph].
%
%
%
\bibitem{pulsars}
  A.~K.~Harding and R.~Ramaty,
  Proc. 20th ICRC, Moscow {\bf 2}, 92-95 (1987); 
%
  A.~M.~Atoian, F.~A.~Aharonian and H.~J.~Volk,
  Phys.\ Rev.\  D {\bf 52} (1995) 3265;
%
  X.~Chi, E.~C.~M.~Young and K.~S.~Cheng,
  Astrophys.\ J.\  {\bf 459} (1995) L83;
%
  C.~Grimani,
  Astron.\ Astrophys.\  {\bf 418}, 649 (2004);
%
  D.~Hooper, P.~Blasi and P.~D.~Serpico,
  arXiv:0810.1527 [astro-ph].
%
%
%
\bibitem{ACR}
See for example  
V.~S.~Berezinskii, S.~V.~Buolanov, V.~A.~Dogiel, 
V.~L.~Ginzburg, V.~S.~Ptuskin,
Astrophysics of Cosmic Rays  (Amsterdam: North--Holland, 1990).
%
%
%
\bibitem{Maurin:2001sj}
  D.~Maurin, F.~Donato, R.~Taillet and P.~Salati,
  Astrophys.\ J.\  {\bf 555} (2001) 585.
%
%
%
\bibitem{Delahaye:2007fr}
  T.~Delahaye, R.~Lineros, F.~Donato, N.~Fornengo and P.~Salati,
  Phys.\ Rev.\  D {\bf 77} (2008) 063527.
%
%
%
\bibitem{Navarro:1995iw}
  J.~F.~Navarro, C.~S.~Frenk and S.~D.~M.~White,
  Astrophys.\ J.\  {\bf 462} (1996) 563.

%
%
%
\bibitem{Bergstrom:1997fj}
  L.~Bergstrom, P.~Ullio and J.~H.~Buckley,
  Astropart.\ Phys.\  {\bf 9} (1998) 137.
%
%
%
\bibitem{Sjostrand:2006za}
  T.~Sj\"ostrand, S.~Mrenna and P.~Skands,
  JHEP {\bf 0605} (2006) 026.
%
%
%
\bibitem{Adriani:2008zq}
  O.~Adriani {\it et al.},
  arXiv:0810.4994 [astro-ph].
%
%
%
\bibitem{egret}
  P.~Sreekumar {\it et al.}  [EGRET Collaboration],
  Astrophys.\ J.\  {\bf 494} (1998) 523.
%
%
%
\bibitem{inpreparation}
In preparation.
%
%
%
\bibitem{FGST}
http://fermi.gsfc.nasa.gov/.
%
%
%
\bibitem{Yin:2008bs}
  P.~f.~Yin, Q.~Yuan, J.~Liu, J.~Zhang, X.~j.~Bi and S.~h.~Zhu,
  arXiv:0811.0176 [hep-ph].
%
%
%
\bibitem{Ishiwata:2008cv}
  K.~Ishiwata, S.~Matsumoto and T.~Moroi,
  arXiv:0811.0250 [hep-ph].

\end{thebibliography}
\end{document}